\documentclass[reprint,onecolumn,amssymb, amsmath,aip,pof]{revtex4-1}

\usepackage{docs}
\usepackage{bm}
\usepackage[colorlinks=true,linkcolor=blue]{hyperref}
\usepackage{color}
\usepackage[usenames,dvipsnames]{xcolor}
\usepackage{graphicx}
\usepackage{dcolumn}
\usepackage[top=2.2cm, bottom=2.2cm, left=3.2cm, right=3.2cm
]{geometry}
\usepackage{subfigure}
\usepackage{amssymb,amsmath}
\usepackage{epstopdf}
\usepackage{wrapfig}
\usepackage{float}
\usepackage{yfonts}
\usepackage[utf8]{inputenc}
\usepackage{enumerate}
\expandafter\ifx\csname package@font\endcsname\relax\else
\expandafter\expandafter
\expandafter\usepackage
\expandafter\expandafter
\expandafter{\csname package@font\endcsname}
\fi
\begin{document}

\preprint{AIP/123-QED}

\newcommand{\mad}[1]{{\textcolor{black}{#1}}}

\newcommand{\md}[1]{{\textcolor{black}{#1}}}
\newcommand{\trp}[1]{{\textcolor{black}{#1}}}

\title{Swimming near Deformable Membranes at Low Reynolds Number}

\author{Marcelo A. Dias }
\affiliation{School of Engineering, Brown University, Providence, Rhode Island 02912}  

\author{Thomas R. Powers}
\affiliation{School of Engineering, Brown University, Providence, Rhode Island 02912}
\affiliation{Department of Physics, Brown University, Providence, Rhode Island 02912}

\date{7 August 2013}

\begin{abstract}

Microorganisms are rarely found in Nature swimming freely in an unbounded fluid. Instead, they typically encounter other organisms, hard walls, or deformable boundaries such as free interfaces or membranes. Hydrodynamic interactions between the swimmer and nearby objects lead to many interesting phenomena, such as changes in swimming speed, tendencies to accumulate or turn, and coordinated flagellar beating. Inspired by this class of problems, we investigate locomotion of microorganisms near deformable boundaries. We calculate the speed of an infinitely long swimmer close to a flexible surface separating two fluids; we also calculate the deformation and swimming speed of the flexible surface. When the viscosities on either side of the flexible interface differ, we find that fluid is pumped along or against the swimming direction, depending on which viscosity is greater.

\end{abstract}


\maketitle

\section{Introduction}

The swimming behavior of motile microorganisms is strongly influenced by the presence of nearby surfaces. Bacteria tend to swim in circles when they swim near a rigid wall~\cite{lauga06}, and hydrodynamic effects influence the accumulation of spermatozoa~\cite{Winet1984} and the swimming behavior of bacteria near rigid walls~\cite{Drescher2011}. Swimmers also encounter deformable surfaces, such as the air-water interface~\cite{Boryshpolets2013}, or the soft mucus-lined tissues in the mammalian female reproductive tract~\cite{Suarez2009}. The deformability of these interfaces helps determine swimming behavior; for example, fish spermatozoa close to an air-water interface have been observed to swim faster than they do at a liquid-water interface~\cite{Boryshpolets2013}. At larger scales, the deformation of the air-water interface is thought to be crucial for the motility of water snails~\cite{Lee2008}. Phenomena such as these have motivated  analytical~\cite{Reynolds1965,Katz1974} and numerical~\cite{Fauci1995} calculations of the swimming behavior of idealized model swimmers at zero Reynolds number (see the references in Lauga and Powers~\cite{Lauga2009}). Perhaps the simplest geometry is a two-dimensional sheet of infinite extent subject to traveling waves of undulation; analytic calculations show that for a deformation of fixed wave speed, wavelength, and amplitude, the presence of a nearby wall makes the swimmer go faster~\cite{Reynolds1965,Katz1974} than it would in the absence of confinement~\cite{Taylor1951}. More recent analytic calculations have explored the effect of walls on propulsive force on an infinite sheet~\cite{Evans2010a}.  Another model is a point-like swimmer such as a force dipole. A deformable interface near such a swimmer exhibiting a reciprocal swimming stroke can lead to net propulsion~\cite{Trouilloud2008} since the interface breaks the symmetry required for the application of the scallop theorem~\cite{Purcell1977}. Finally, numerical calculations for a swimming sheet in a viscoelastic medium near an elastic membrane have shown that nearby elastic structures can enhance both the swimming speed and efficiency~\cite{Chrispell2013}. Likewise, numerical calculations of a model dipole swimmer in an elastic tube also show an increase in swimming speed due to the elasticity of the walls~\cite{Ledesma-Aguilar2013}. In this article we calculate the swimming speed for a swimming sheet near an elastic membrane in the limit of small-amplitude waves. Our work is  related to that of Chrispell \emph{et al.}~\cite{Chrispell2013} but is distinct from it in several ways. Although we give explicit analytic formulas for the swimming speed for any membrane tension $\gamma$ and bending stiffness $B$, we focus on the limit of vanishing tension, and study how the interplay of bending and viscous effects determines the speed. Second, we pay special attention to the induced swimming of the elastic sheet, and illustrate how the direction is determined by the superposition of transverse and longitudinal deformations of the membrane. And third, we allow the viscosities on either side of the membrane to differ, and find that when the viscosities are unequal, the swimmer drags fluid along or against the swimming direction, depending on which viscosity is larger. \md{This phenomenon is distinct from transport of fluid in ciliated tubes of mammalian reproductive systems~\cite{Brennen1977}, or peristaltic pumping, where two deforming surfaces are prevented from translating by an external forces, thus causing transport of fluid along the channel~\cite{Felderhof2009}.}

\section{Description of the model}

\begin{figure}[h]
    \centering
    \includegraphics[width=5in]{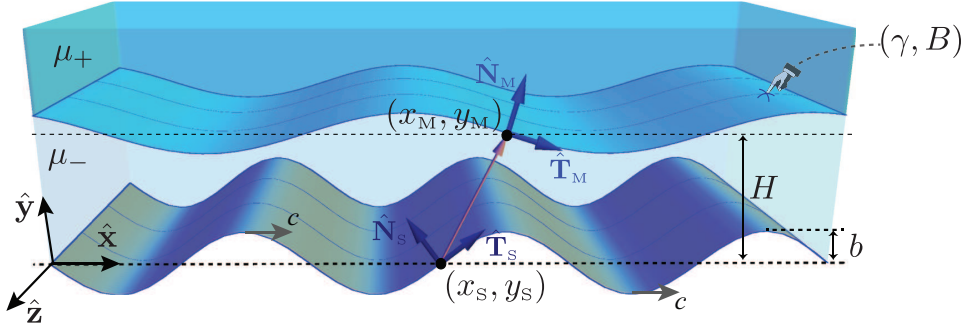}
    \caption{(Color online) Infinite swimming sheet of amplitude $b$ beneath a membrane at average height $H$, which separates two fluids with viscosities $\mu_{{\scriptscriptstyle\pm}}$. The membrane has surface tension $\gamma$ and bending rigidity $B$. The coordinates $(x_{\mbox{{\tiny S}}},y_{\mbox{{\tiny S}}})$ and $(x_{\mbox{{\tiny M}}},y_{\mbox{{\tiny M}}})$ describe the swimmer and interface, respectively.}
    \label{fig:SwimmerModel}
\end{figure}

Figure~\ref{fig:SwimmerModel} shows \mad{the setup of} our model problem: a membrane with surface tension $\gamma$ and bending rigidity $B$ lies near a swimming sheet. The average separation between the sheet and the membrane is $H$, and the membrane separates two fluids with different viscosities,  $\mu_{{\scriptscriptstyle\pm}}$. We disregard the space under the swimming sheet\mad{, since including a third fluid region introduces more complexity without adding any fundamentally new phenomena.} \mad{We work in the frame of the swimmer.} The material points on the swimming sheet have coordinates $(x_{\mbox{{\tiny S}}},y_{\mbox{{\tiny S}}})$, and the deformation of the swimmer is a prescribed transverse wave, with $x_{\mbox{{\tiny S}}}(x,t)=x$ and $y_{\mbox{{\tiny S}}}(x,t)=b \sin\left[k(x-ct)\right]$. The amplitude $b$ of the wave is assumed much smaller than the wavelength: $b\,k\ll1$. Since $b$ is treated as infinitesimal, it is also always small compared to $H$.  The material points on the sheet 
have velocity components given by $u_{\mbox{{\tiny S}}}={\partial x_{\mbox{{\tiny S}}}}/{\partial t}=0$ and $v_{\mbox{{\tiny S}}}={\partial y_{\mbox{{\tiny S}}}}/{\partial t}=-c\,kb \cos\left[k(x-ct)\right]$.

Given the deformation of the swimmer, the problem is to find the the velocity flow everywhere, the swimming speed of the sheet, and  the shape and swimming speed of the membrane. Following Taylor~\cite{Taylor1951}, we expand in powers of $b\,k$ and find the swimming speed to leading order in $b\,k$; just as in Taylor's case, the symmetry of the problem under $b\mapsto-b$ makes the swimming speed an even function of $b\,k$.  The governing equations for the fluid are Stokes equations, $-\boldsymbol\nabla p_{{\scriptscriptstyle\pm}}+\mu_{{\scriptscriptstyle\pm}}\nabla^2\mathbf{v}_{{\scriptscriptstyle\pm}}=\bm{0}$ and $\boldsymbol\nabla \cdot\mathbf{v}_{{\scriptscriptstyle\pm}}=0$, where the $\pm$ subscript denotes the region above or below the membrane. The problem is two-dimensional since there is no dependence on $z$. The shape of the passive membrane is given by $(x_{\mbox{{\tiny M}}},y_{\mbox{{\tiny M}}})$, which are unknown functions of $x$ and $t$.  At the surface of the swimmer and the membrane we impose no-slip boundary conditions,
\begin{subequations}
\label{eq:BCSwimmerD}
	\begin{align}
		\label{eq:BCSwimmerD1}
			&\left.(v_x^-,v_y^-)\right|_{{x}_{\mbox{{\tiny S}}},{y}_{\mbox{{\tiny S}}}}=(u_{\mbox{{\tiny S}}},v_{\mbox{{\tiny S}}})
			\\
		\label{eq:BCInterfaceD1}
			&\left.(v_x^-,v_y^-)\right|_{{x}_{\mbox{{\tiny M}}},{y}_{\mbox{{\tiny M}}}}=\left.(v_x^+,v_y^+)\right|_{{x}_{\mbox{{\tiny M}}},{y}_{\mbox{{\tiny M}}}}			\\
		\label{eq:BCInterfaceD2}
			&\left.(v_x^-,v_y^-)\right|_{{x}_{\mbox{{\tiny M}}},{y}_{\mbox{{\tiny M}}}}=(u_{\mbox{{\tiny M}}},v_{\mbox{{\tiny M}}})\equiv\left(\left.\frac{\partial x_{\mbox{{\tiny M}}}}{\partial t}\right|_x,\left.\frac{\partial y_{\mbox{{\tiny M}}}}{\partial t}\right|_x\right).
	\end{align}
\end{subequations}
Note that $x$ serves as a Lagrangian label for the swimmer and the membrane, with $x_{\mbox{{\tiny M}}}(x,t=0)=x$ when $b=0$.

\md{Additional boundary conditions are required to determine the flow velocity and the deformation of the membrane. The balance of  external viscous traction and internal elastic forces per unit area determines the shape of the membrane:
\begin{subequations}
	\label{eq:ForceBalanceInterfaceD}
		\begin{align}
			\label{eq:ForceBalanceInterfaceD2}
				&\left.\hat{\mathbf{N}}_{\mbox{{\tiny M}}}\cdot\left(\boldsymbol\sigma_{{\scriptscriptstyle+}}-\boldsymbol\sigma_{{\scriptscriptstyle-}}\right)\cdot\hat{\mathbf{N}}_{\mbox{{\tiny M}}}\right|_{x_{\mbox{{\tiny M}}},y_{\mbox{{\tiny M}}}}=\left[-\gamma\kappa+B\left(\kappa''+\frac{1}{2}\kappa^3\right)\right]_{x_{\mbox{{\tiny M}}}}\\
			\label{eq:ForceBalanceInterfaceD3}
				&\left.\hat{\mathbf{T}}_{\mbox{{\tiny M}}}\cdot\left(\boldsymbol\sigma_{{\scriptscriptstyle+}}-\boldsymbol\sigma_{{\scriptscriptstyle-}}\right)\cdot\hat{\mathbf{N}}_{\mbox{{\tiny M}}}\right|_{x_{\mbox{{\tiny M}}},y_{\mbox{{\tiny M}}}}=0,
		\end{align}
\end{subequations} 
where $\kappa''$ is the second derivative of the curvature with respect to arclength~\cite{Powers2010}. 
 In Eq.~\eqref{eq:ForceBalanceInterfaceD}, $\hat{\mathbf{N}}_{\mbox{{\tiny M}}}$ is the upward-pointing unit normal of the membrane, $\hat{\mathbf{T}}_{\mbox{{\tiny M}}}$ is the unit tangent vector to the membrane, $\boldsymbol\sigma_{{\scriptscriptstyle\pm}}=-p_{{\scriptscriptstyle\pm}}\mathbb{I}+\mu_{{\scriptscriptstyle\pm}}\left[\boldsymbol\nabla\mathbf{v}_{{\scriptscriptstyle\pm}}+\left(\boldsymbol\nabla\mathbf{v}_{{\scriptscriptstyle\pm}}\right)^T\right]$ is the viscous stress with $p_{{\scriptscriptstyle\pm}}$ the pressure and $ \mathbb{I}$ is the identity matrix. Note that the tangential stress in Eq.~(\ref{eq:ForceBalanceInterfaceD3}) vanishes since we assume the tension $\gamma$ is uniform~\cite{Powers2010}.} 
 
\mad{The force balance equations~\eqref{eq:ForceBalanceInterfaceD} imply that the $x$-component of force $\langle{\mathbf{f}}_{\mbox{{\tiny M}}}\rangle_x$ on the membrane is zero: 
\begin{eqnarray}
	\langle{\mathbf{f}}_{\mbox{{\tiny M}}}\rangle_x&\equiv&\frac{1}{L}\int_0^{L}\mathrm{d}s\,\hat{\mathbf{x}}\cdot\left(\boldsymbol\sigma_{{\scriptscriptstyle+}}-\boldsymbol\sigma_{{\scriptscriptstyle-}}\right)\cdot\hat{\mathbf{N}}_{\mbox{{\tiny M}}}\nonumber\\
	&=&\frac{1}{L}\int_0^{L}\mathrm{d}s\,\hat{\mathbf{x}}\cdot\hat{\mathbf{N}}_{\mbox{{\tiny M}}}\left[-\gamma\kappa+B\left(\kappa''+\frac{1}{2}\kappa^3\right)\right]_{x_{\mbox{{\tiny M}}}}\nonumber\\
	&=&\frac{1}{L}\hat{\mathbf{x}}\cdot\int_0^{L}\mathrm{d}s\,\frac{\mathrm{d}}{\mathrm{d}s}\left[-\gamma\hat{\mathbf{T}}_{\mbox{{\tiny M}}}-B\left(\frac{\kappa^2}{2}\hat{\mathbf{T}}_{\mbox{{\tiny M}}}+\kappa^\prime\hat{\mathbf{N}}_{\mbox{{\tiny M}}}\right)\right]=0.
\end{eqnarray}
As we shall see below, 
the membrane translates in the $x$ direction even though the $x$-component of force vanishes; therefore, the membrane behaves as a passive swimmer.}
 
Since the Reynolds number vanishes, the net force on the \mad{active} swimmer vanishes, $\mathbf{F}=\int_{\mbox{{\tiny S}}}{\boldsymbol\sigma}_{{\scriptscriptstyle-}}\cdot\hat{\mathbf{N}}_{\mbox{{\tiny S}}}\,\mathrm{d}A=0$. In particular, the $x$-component of the hydrodynamic force  per wavelength acting on the swimmer must also vanish:
\begin{equation}
	\label{eq:AverageForceSheetD}
		\langle{\mathbf{f}}_{\mbox{{\tiny S}}}\rangle_x\equiv\frac{k}{2\pi}\int_0^{2\pi/k}\mathrm{d}x\left.\hat{\mathbf{x}}\cdot{\boldsymbol\sigma}_{{\scriptscriptstyle-}}\cdot\hat{\mathbf{N}}_{\mbox{{\tiny S}}}\right|_{{x}_{\mbox{{\tiny S}}},{y}_{\mbox{{\tiny S}}}}=0.
\end{equation}
We will use the notation $\langle\cdot\rangle$ to denote the average over a wavelength. We have listed all the boundary conditions required to find the flow velocity and the shape of the membrane. Note that if the deformation of the sheet leads to an average flow $\langle \hat{\mathbf{x}}\cdot\mathbf{v}_{{\scriptscriptstyle+}}(y\rightarrow\infty)\rangle$, then the sheet swims to the left with speed $V_{\mbox{{\tiny S}}}=\langle \hat{\mathbf{x}}\cdot\mathbf{v}_{{\scriptscriptstyle+}}(y\rightarrow\infty)\rangle$ in the laboratory frame. In the swimmer frame, the average speed of the membrane is ${\langle u_{\mbox{{\tiny M}}}\rangle}=(k/2\pi)\int_0^{2\pi/k}\mathrm{d}x\,\left.\partial{\psi}_{{\scriptscriptstyle\pm}}/\partial {y}\right|_{{x}_{\mbox{{\tiny M}}},{y}_{\mbox{{\tiny M}}}}$.
\noindent Therefore,  the membrane swims to the left with speed $V_{\mbox{{\tiny M}}}=V_{\mbox{{\tiny S}}}-\langle u_{\mbox{{\tiny M}}}\rangle$ in the laboratory frame.

\section{Perturbative expansion}

Before solving the governing equations subject to the boundary conditions, it is convenient to cast them in dimensionless form by measuring length in units of $1/k$, rates in units of $c\,k$, and stress in units of $c\,k\,\mu_{{\scriptscriptstyle-}}$. These choices lead to the following natural dimensionless groups: the viscosity ratio  $\mu_r\equiv\mu_{{\scriptscriptstyle+}}/\mu_{{\scriptscriptstyle-}}$, the capillary number ${C\!a}\equiv c\,\mu_{{\scriptscriptstyle-}}/\gamma$, and the Machin number  ${M\!a}\equiv k^2B/\left(c\,\mu_{{\scriptscriptstyle-}}\right)$. Since the flow is incompressible and two-dimensional, it is also natural to work in terms of the stream function $\psi_{{\scriptscriptstyle\pm}}$, where $\mathbf{v}_{{\scriptscriptstyle\pm}}={\boldsymbol\nabla}\times(\hat{\mathbf z}\,\psi_{{\scriptscriptstyle\pm}})$. Taking the curl of the Stokes equations reveals that $\psi_{{\scriptscriptstyle\pm}}$ is biharmonic, $\nabla^2\psi_{{\scriptscriptstyle\pm}}=0$. Since we must simultaneously solve for the membrane shape using the force-balance conditions of Eq.~\eqref{eq:ForceBalanceInterfaceD2}, we also solve for the pressure.

Expanding in powers of $b\,k$, and introducing the convenient variable $\zeta=x-t$, we have
\begin{subequations}
	\label{eq:UnknownsExpand}
		\begin{align}
			\label{eq:UnknownsExpand1}
				(\psi_{{\scriptscriptstyle\pm}},p_{{\scriptscriptstyle\pm}})&=\sum_{n=1}^\infty (b\,k)^n \left(\Psi^{(n)}_{{\scriptscriptstyle\pm}}(\zeta,y),P^{(n)}_{{\scriptscriptstyle\pm}}(\zeta,y)\right)\\
			\label{eq:UnknownsExpand3}
				(x_{\mbox{{\tiny M}}},y_{\mbox{{\tiny M}}})&=\left(x+t\,\langle u_{\mbox{{\tiny M}}}\rangle+\sum_{n=1}^\infty (b\,k)^n X^{(n)}(\zeta),
			H+\sum_{n=1}^\infty (b\,k)^n Y^{(n)}(\zeta)\right),
		\end{align}
\end{subequations}
where $\langle u_{\mbox{{\tiny M}}}\rangle=\sum_{n=1}^\infty (b\,k)^n U_{\mbox{{\tiny M}}}^{(n)}$. Because we work in the frame of the swimming organism, we require that $\partial \psi^{(2)}/\partial y$ approaches a constant and $\partial \psi^{(2)}/\partial x$ approaches zero as $y\rightarrow \infty$. Under these conditions, the general solution for the biharmonic equation is
\begin{subequations}
	\label{eq:SolBiharmonic}
		\begin{align}
			\label{eq:SolBiharmonic1}
				\Psi ^{(n)}_{{\scriptscriptstyle-}}(\zeta,y)&=\!\!\sum _{m=1}^n \!\!\left\{\left[\left(yA^{(m,n)}_{{\scriptscriptstyle-}}\!\!+\!B^{(m,n)}_{{\scriptscriptstyle-}}\right) \sin (m\zeta)\!+\!\left(yC^{(m,n)}_{{\scriptscriptstyle-}}\!\!+\!D ^{(m,n)}_{{\scriptscriptstyle-}}\right)\cos (m\zeta)\right] \sinh (my)\right.\nonumber\\
				&\left. +\!\!\left[\!\left(\!yE^{(m,n)}_{{\scriptscriptstyle-}}\!\!+\!F^{(m,n)}_{{\scriptscriptstyle-}}\!\!\right) \!\sin (m\zeta)\!+\!\left(\!y G^{(m,n)}_{{\scriptscriptstyle-}}\!\!+\!H^{(m,n)}_{{\scriptscriptstyle-}}\!\!\right)\! \cos (m\zeta)\!\right]\! \cosh (my)\!\right\}\!\nonumber\\
				&+\! yu^{(1,n)}_{{\scriptscriptstyle-}}\!\!+\!y^2 u^{(2,n)}_{{\scriptscriptstyle-}}\!\!+\!y^3u^{(3,n)}_{{\scriptscriptstyle-}}\\
			\label{eq:SolBiharmonic2}
				\Psi ^{(n)}_{{\scriptscriptstyle+}}(\zeta,y)&=\!\!\sum _{m=1}^n\!\! \left[\left(\!yA^{(m,n)}_{{\scriptscriptstyle+}}\!\!+\!B^{(m,n)}_{{\scriptscriptstyle+}}\!\!\right)\! \sin (m\zeta)\!+\!\left(\!yC^{(m,n)}_{{\scriptscriptstyle+}}\!\!+\!D ^{(m,n)}_{{\scriptscriptstyle+}}\!\!\right)\!\cos (m\zeta)\!\right] \!e^{-my}\!\!+\! yu^{(1,n)}_{{\scriptscriptstyle+}}\!\!,
		\end{align}
\end{subequations}
where $\left\{A^{(m,n)}_{{\scriptscriptstyle\pm}},B^{(m,n)}_{{\scriptscriptstyle\pm}},C^{(m,n)}_{{\scriptscriptstyle\pm}},D^{(m,n)}_{{\scriptscriptstyle\pm}},E^{(m,n)}_{{\scriptscriptstyle-}},F^{(m,n)}_{{\scriptscriptstyle-}},G^{(m,n)}_{{\scriptscriptstyle-}},H^{(m,n)}_{{\scriptscriptstyle-}},u^{(1,n)}_{{\scriptscriptstyle\pm}},u^{(2,n)}_{{\scriptscriptstyle-}},u^{(3,n)}_{{\scriptscriptstyle-}}\right\}$ are functions of the dimensionless parameters $\{{C\!a}, {M\!a}, \mu_r, H\}$  to be determined using the boundary conditions, Eqs.~\eqref{eq:BCSwimmerD}--\eqref{eq:AverageForceSheetD}. 
A nonzero $u^{(1,n)}_{{\scriptscriptstyle-}}$ leads to a uniform flow in the region between the swimmer and the membrane, $u^{(2,n)}_{{\scriptscriptstyle-}}$ leads to pure shear flow,  $u^{(3,n)}_{{\scriptscriptstyle-}}$ leads to parabolic flow,  and $u^{(1,n)}_{{\scriptscriptstyle+}}$ leads to an average velocity of the fluid at infinity in the swimmer's frame.

\md{Note that the nonlinearity of the problem enters through the boundary conditions, which must be expanded to second order using the expressions in Eq..~\ref{eq:UnknownsExpand3}. Furthermore, we use the small-slope approximation to write  $\kappa\approx-\partial^2 y_{\mbox{{\tiny M}}}/\partial x^2$ and $\kappa''\approx-\partial^4 y_{\mbox{{\tiny M}}}/\partial x^4 $. We need only retain the $\mathcal{O}(b\,k)$ part of $\kappa$  and the bending force per unit area in Eq.~\eqref{eq:ForceBalanceInterfaceD2} because the corrections are $\mathcal{O}(b^3k^3)$ and the swimming velocity turns out to be $\mathcal{O}(b^2k^2)$.}
\md{The first-order and second-order results are shown in appendices~\ref{SIB} and~\ref{SIC}, respectively.}


\section{Results}

\begin{figure}[h]
    \centering
    \includegraphics[width=5.5in]{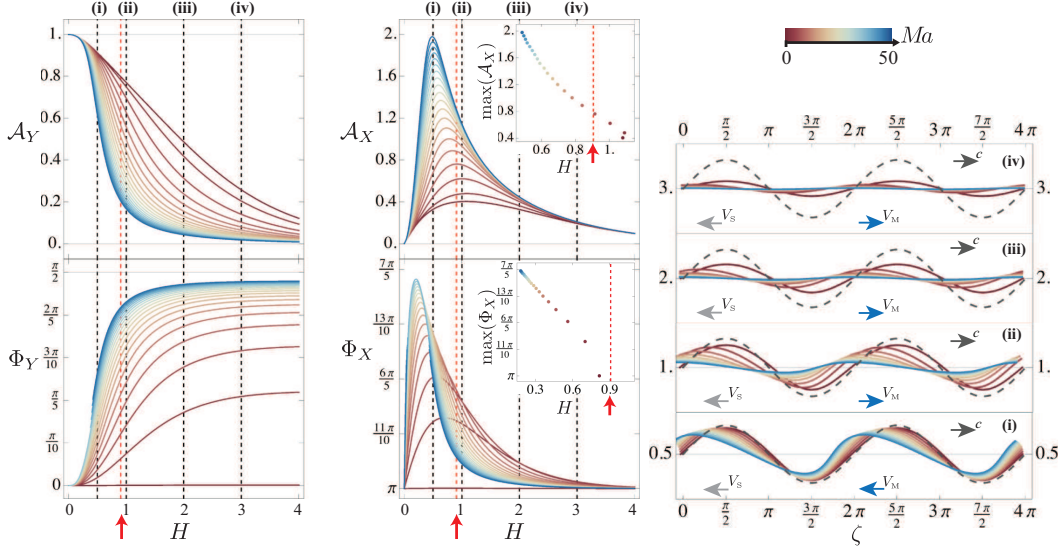}
    \caption{(Color online) Amplitudes and phases of the transverse (left panel) and longitudinal (middle panel) waves of the membrane. The insets for the longitudinal waves show the maximum values. In the right panel, membrane shapes are shown for various heights, with $b=0.3$, and the shape of the swimmer (gray dashed line) is superimposed. The directions of the swimmer wave speed, swimmer swimming speed, and membrane swimming speed are also indicated with arrows. Here $C\!a\rightarrow\infty$ and $\mu_r=1/2$. The color scheme is a function of the Machin number $M\!a$. The dashed vertical red lines (see vertical arrows) indicate the height, $H\approx0.91$, at which the membrane reverses its direction. The other vertical lines in the left and middle panels correspond to (i) $H=1/2$, (ii) $H=1$, (iii) $H=2$, and (iv) $H=3$.}
    \label{fig:InterfaceShape}
\end{figure}

As mentioned previously, there is no swimming velocity to first order in $b\,k$. However, the flow induced by the motion of the swimmer causes the membrane to have a first-order deformation. Solving (\ref{eq:BCInterfaceD2}) and (\ref{eq:ForceBalanceInterfaceD2}) yields (see Appendix \ref{SIB})
\begin{subequations}
	\label{eq:ShapeSolM}
		\begin{align}
			\label{eq:ShapeSolMX}
				x_{\mbox{{\tiny M}}}&=x+b\,k\,\mathcal{A}_{X}\cos\left(\zeta-\Phi_{X}\right)+\mathcal{O}(b^2)\\
			\label{eq:ShapeSolMY}
				y_{\mbox{{\tiny M}}}&=H+b\,k\,\mathcal{A}_{Y}\sin\left(\zeta+\Phi_{Y}\right)+\mathcal{O}(b^2),
		\end{align}
\end{subequations} 
where the dependence of the amplitudes $\{\mathcal{A}_{X},\mathcal{A}_{Y}\}$ and phases $\{\Phi_{X},\Phi_{Y}\}$ on $M\!a$ and $H$ is shown in Fig.~\ref{fig:InterfaceShape} (left and middle panels) for the case of $C\!a\rightarrow\infty$ and $\mu_r=\mu_{{\scriptscriptstyle+}}/\mu_{{\scriptscriptstyle-}}=1/2$. Figure~\ref{fig:InterfaceShape} (right panel) also displays the shape of the membrane for various $M\!a$ for a few different values of $H$.

When the membrane is close to the swimmer, the waves induced on the membrane are mostly transverse, with an amplitude and phase close to that of the swimmer. As the separation $H$ increases, the transverse wave amplitude $\mathcal{A}_Y$ decreases while the longitudinal wave amplitude $\mathcal{A}_X$ increases, reaching a maximum at a value of $H$ that depends on $M\!a$. Eventually $\mathcal{A}_X$ also decreases with $H$ since the hydrodynamic interaction between the swimmer and membrane decreases with distance. Since there is no force preventing the membrane from translating, the induced waves cause the membrane to swim. We will see below that when the separation is small and transverse waves dominate, the membrane swims in the same direction as the swimmer. When the separation is large and longitudinal waves dominate, the membrane swims in the opposite direction as the swimmer, which is to be expected since sheet with purely longitudinal waves swims in the opposite direction as a sheet with purely transverse waves. We will see below that this explanation for swimming direction does not quantitatively predict the reversal in swimming direction of the membrane since it disregards the interplay of the phases of the wave, and since it does not completely account for the interaction between the membrane and the swimmer. 

As validation, we have checked (see Appendix \ref{SID}) that our first-order stream function with ${M\!a}\rightarrow0$, ${C\!a}\rightarrow\infty$, and $\mu_r\rightarrow1$ agrees with Taylor's result for swimming in an unbounded fluid~\cite{Taylor1951}, and that the stream function with ${M\!a}\rightarrow\infty$, ${C\!a}\rightarrow0$, and $\mu_r\rightarrow\infty$ agrees with Reynolds'~\cite{Reynolds1965} and Katz's~\cite{Katz1974} results for swimming near a rigid wall.

We now turn to the second-order calculation (see Appendix \ref{SIC}). Since we only seek time-averaged properties, we need only compute the unknowns $\left\{u^{(1,2)}_{{\scriptscriptstyle\pm}},u^{(2,2)}_{{\scriptscriptstyle-}},u^{(3,2)}_{{\scriptscriptstyle-}}\right\}$ and the average interface speed $U_{\mbox{{\tiny M}}}^{(2)}$. We impose the following boundary conditions: (i) the average of the no-slip condition over one spatial period (\ref{eq:BCSwimmerD1}); (ii) the tangential force-balance condition on the membrane (\ref{eq:ForceBalanceInterfaceD3}); (iii) the force-free condition on the swimmer (\ref{eq:AverageForceSheetD}); and the no-slip condition at the membrane, which amounts to two conditions, (iv) below and (v) above the membrane. These five conditions are sufficient to determine the five unknowns.

\begin{figure}[h]
    \centering
    \includegraphics[width=5.5in]{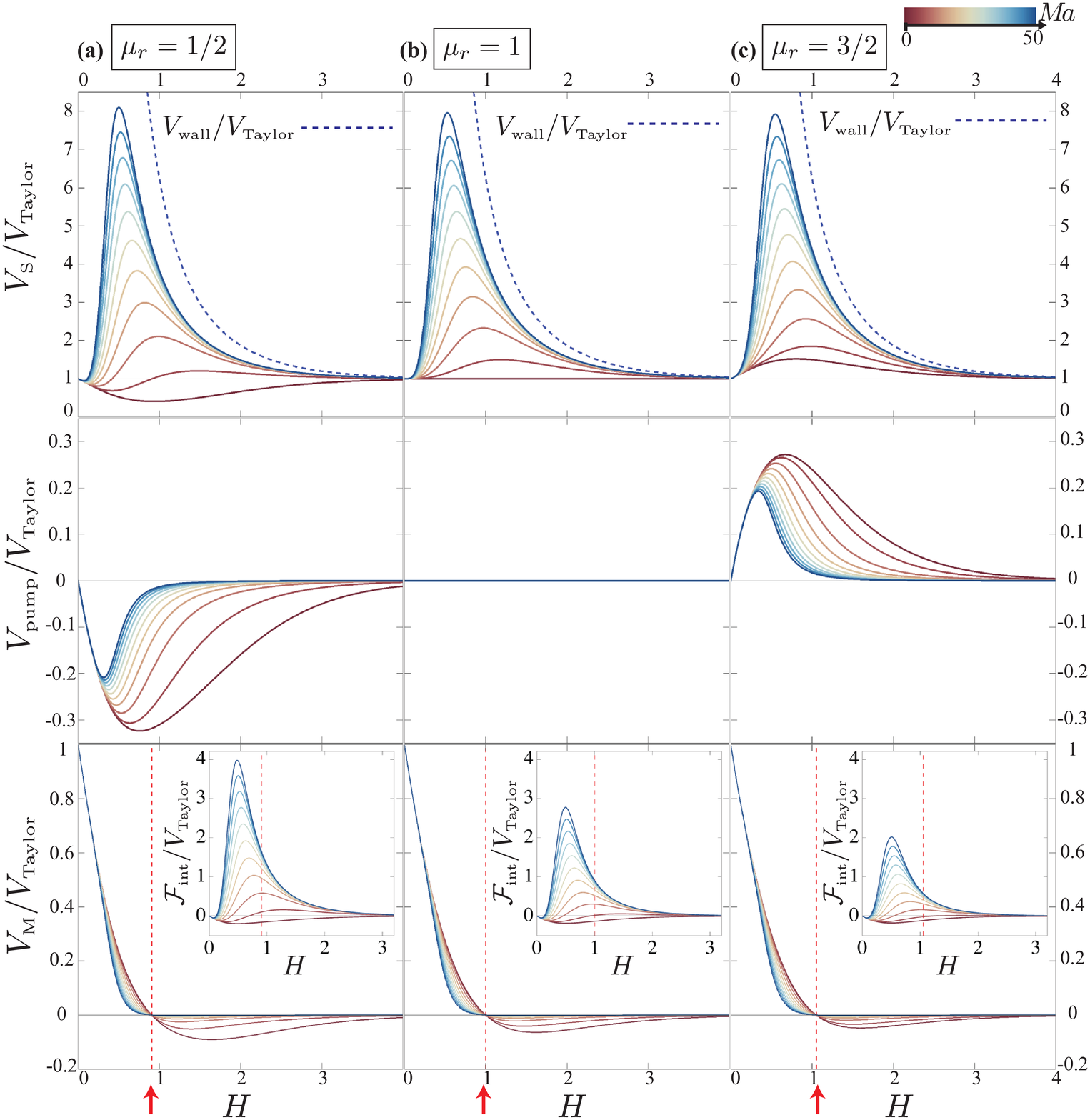}
    \caption{(Color online) Speeds normalized by the Taylor result in the laboratory frame, where $V_{\mbox{\tiny Taylor}}=cb^2k^2/2$. These results are in the limit where ${C\!a}\rightarrow\infty$, for three different values of viscosity ratio ($\mu_r=1/2$, $\mu_r=1$, and $\mu_r=3/2$) and the range of $M\!a\in[0,50]$. The insets in the bottom row show the behavior of  $\mathcal{F}_{\mbox{\tiny int}}/V_{\mbox{\tiny Taylor}}$ as a function of $H$. The vertical dashed lines (see vertical arrows) in the bottom row and in the insets indicate the separation $H$ at which the membrane speed reverses.}
    \label{fig:Speeds2}
\end{figure} 

\md{For simplicity, here we consider a geometry in which either the flow beneath the swimmer is disregarded or the swimmer is moving in a symmetric channel. This simplification allows us to eliminate the contribution coming from pure shear, $u^{(2,2)}_{{\scriptscriptstyle-}}$. For an asymmetric channel, we must consider the flow below the swimmer, and the only other change is that the force-free condition on the swimmer will involve the fluid stresses from both sides of the swimmer, which leads to a nonzero shear flow. The force-free condition makes $u^{(3,2)}_{{\scriptscriptstyle-}}$ vanish whether or not the channel is symmetric.} Translating back to the laboratory frame by subtracting off the average flow velocity at $y\rightarrow\infty$ yields (in dimensional form) the average fluid velocity within the gap between the swimmer and the membrane $V_{\mbox{\tiny pump}}=cb^2k^2\left(u^{(1,2)}_{{\scriptscriptstyle+}}- u^{(1,2)}_{{\scriptscriptstyle-}}\right)$, the membrane swimming speed $V_{\mbox{{\tiny M}}}=cb^2k^2\left(u^{(1,2)}_{{\scriptscriptstyle+}}- U_{\mbox{{\tiny M}}}^{(2)}\right)$, and the speed of the swimmer, $V_{\mbox{{\tiny S}}}=cb^2k^2\left(u^{(1,2)}_{{\scriptscriptstyle+}}-0\right)$. These expressions are too complicated to display here, but they simplify in the case of equal viscosities, $\mu_r=1$, for which we find $V_{\mbox{\tiny pump}}=0$, and 
\begin{subequations}
\begin{align}
V_{\mbox{\tiny S}}&=cb^2k^2\left[\frac{1}{2}-\frac{2 H (1+H) \left[1-e^{2 H}+2 H (1+H)\right]}{e^{4 H}\Omega^2+\left[1-e^{2 H}+2 H (1+H)\right]^2}\right]\\
V_{\mbox{\tiny M}}&= \frac{cb^2k^2}{2}\frac{e^{2 H}\left(1-H^2\right)\Omega^2}{e^{4 H}\Omega^2+\left[1-e^{2 H}+2 H (1+H)\right]^2},\label{eq:VM}
\end{align}
\end{subequations}
%
%
where $\Omega\equiv4C\!a/(1+C\!aM\!a)$. Equation~\eqref{eq:VM} shows that the membrane speed vanishes at $H=1$ when $\mu_r=1$. From Fig.~\ref{fig:Speeds2} we can see that the membrane speed vanishes at $H>1$ when $\mu_r>1$, and at $H<1$ when $\mu_r<1$ (indicated by the red dashed line). Even in the case of unequal viscosities we find that the swimming velocities depend on the material properties through $\Omega$. Thus, \mad{the form of} $\Omega$ shows how the swimming speeds in the limit of zero capillary number and infinite Machin number are the same.

We now concentrate on the effects of bending rigidity by taking the limit ${C\!a}\rightarrow\infty$. The plots in Fig.~\ref{fig:Speeds2} show the swimmer, pumping, and membrane speeds as a function of $H$, for the range $M\!a\in[0,50]$ and $\mu_r=1/2$, $\mu_r=1$, and $\mu_r=3/2$. First note that we recover Taylor's results~\cite{Taylor1951} when ${M\!a}\rightarrow0$ and $\mu_r=1$; in this limit we find the swimming speed is independent of $H$ and there is no net flux of fluid entrained by the swimmer (Fig.~\ref{fig:Speeds2}, middle column). For all viscosity ratios, increasing the bending stiffness leads to enhancement of the swimmer's speed and diminishment of the pumping and membrane speeds. The sign of the pumping speed is determined by the viscosity ratio alone, and is negative if $\mu_r<1$, zero if $\mu_r=1$, and positive if $\mu_r>1$. The swimmer develops its maximum speed in the limit where ${M\!a}\rightarrow\infty$, ${C\!a}\rightarrow0$, and $\mu_r\rightarrow\infty$. This limit, shown in Fig.~\ref{fig:Speeds2} as the dashed blue curve, is the hard-wall result derived previously in the literature~\cite{Reynolds1965,Katz1974}. As mentioned previously, the sign of the membrane speed determined from the first-order results (\ref{eq:ShapeSolM}). A swimmer in an unbounded fluid~\cite{Blake1971} or in the presence of a wall~\cite{Katz1974} can swim backwards or forwards depending upon the competition between the longitudinal and transverse waves. Our results are also in agreement with this tendency:
\begin{equation}
		V_{\mbox{{\tiny M}}}=\frac{cb^2k^2}{2}\left[\mathcal{A}_Y^2+\mathcal{A}_Y\mathcal{A}_X\cos\left(\Phi_X+\Phi_Y\right)-\mathcal{A}_X^2\right]+\mathcal{F}_{\mbox{\tiny int}},
\end{equation}
where the new term $\mathcal{F}_{\mbox{\tiny int}}$ is a correction due to the elastic properties of the interface and the hydrodynamic interaction between the swimmer and the membrane. The dependence of $\mathcal{F}_{\mbox{\tiny int}}$ is shown in the insets in the third row of Fig.~\ref{fig:Speeds2}.  Note that as  $\mu_r$ gets larger, the amplitude of $\mathcal{F}_{\mbox{\tiny int}}$ decreases. For most values of $M\!a$, the peak of $\mathcal{F}_{\mbox{\tiny int}}$ lies to the left of the red dashed line. Therefore, this interaction tends to help the membrane swim in the same direction of the swimmer. The height in which the membrane speed changes sign, as shown in Fig.~\ref{fig:Speeds2}, depends on the viscosity ratio. Comparing the results in Figs.~\ref{fig:InterfaceShape} and \ref{fig:Speeds2}, we observe that positive membrane speeds are below the threshold (marked by the dashed, red, vertical line in Fig.~\ref{fig:InterfaceShape}) below which the maxima of $\Phi_X$ are located.

\begin{figure}[h]
    \centering
    \includegraphics[width=5.5in]{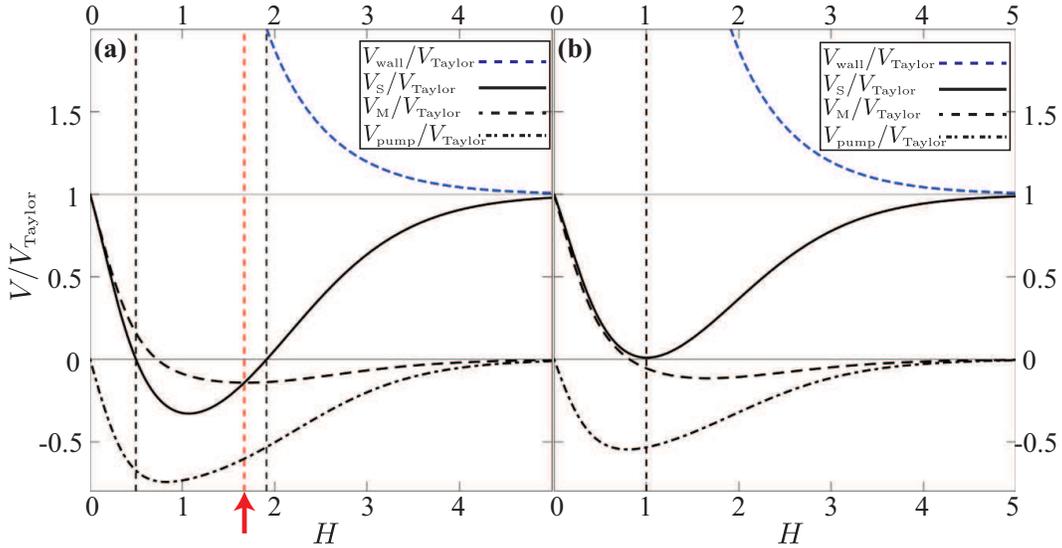}
    \caption{(Color online) Speeds normalized by the Taylor result in the laboratory frame, where $V_{\mbox{\tiny Taylor}}=cb^2k^2/2$. These results are in the limit where ${M\!a}\rightarrow0$ and ${C\!a}\rightarrow\infty$. In {(a)} $\mu_r\rightarrow0$ and {(b)} $\mu_r\approx0.215$.}
    \label{fig:Speeds1}
\end{figure}

Figure~\ref{fig:Speeds1} shows the speeds as a function of the membrane's average height, in the limit ${M\!a}\rightarrow0$ and ${C\!a}\rightarrow\infty$, for two special values of the viscosity ratio,  {(a)} $\mu_r\rightarrow0$ and {(b)} $\mu_r\approx0.215$. These plots show that for a small enough viscosity ratio there is always a regime in which the swimmer has negative speed. In Fig.~\ref{fig:Speeds1}~{(a)}, where $\mu_{{\scriptscriptstyle-}}/\mu_{{\scriptscriptstyle+}}\rightarrow\infty$, the swimmer has negative speed (swims in the direction of the propagating waves on the sheet) when $H\in[0.5,1.9]$ (dashed black vertical lines). Below $H\approx1.7$ (dashed red vertical line indicated by the arrow), the membrane swims faster than the swimmer. The critical value of $\mu_r$ is $\mu_r\approx0.215$ where the swimmer's speed vanishes at $H=1$ (Fig.~\ref{fig:Speeds1}~{(b)}). It is a general feature of this system that whenever $\mu_r<1$, the pumping speed is always negative. The pumping is always positive when $\mu_r>1$.

\section{Concluding remarks}

To summarize, we have studied the problem of a swimming sheet near a deformable membrane with bending rigidity and a constant surface tension. Our main results are the swimming speed of the sheet as well as the deformation and swimming speed of the membrane.  A distinctive feature of our problem that does not occur in the problem of a sheet swimming in unbounded fluid is that fluid is dragged along with or propelled backwards past the swimmer. Future work should lift some of the simplifying assumptions we have made, and examine finite-length swimmers with large amplitude deformations near flexible surfaces. Another generalization would be to consider swimming near an elastic half-space.

\acknowledgments
This work was supported in part by National Science Foundation Grant No. CBET-0854108. We are grateful to D. T.-N. Chen, M. S. Krieger, A. Morozov, and S. Spagnolie for discussions and comments on the manuscript.

\appendix
\section{Dimensionless quantities}
\label{SIA}

We start by conveniently expressing our system of equations and boundary conditions in terms of dimensionless quantities. In cases which $H\,k\sim\mathcal{O}(1)$, but $b\,k\ll\mathcal{O}(1)$, we choose the wavelength $2\pi/k$ as the appropriate length scale of the problem. The choice of time comes from the beating frequency of the wave, $c\,k$. Therefore, we may define the following dimensionless variables: $\tilde{x}\equiv k\,x$, $\tilde{y}\equiv k\,y$, $\tilde{b}\equiv k\, b$, $\tilde{H}\equiv k\, H$, $\tilde{t}\equiv c\,k\, t$, $\tilde{\langle u_{\mbox{{\tiny M}}}\rangle}\equiv \langle u_{\mbox{{\tiny M}}}\rangle/c$, and $\tilde{p}_{{\scriptscriptstyle\pm}}\equiv p_{{\scriptscriptstyle\pm}}/(c\,k\,\mu_{{\scriptscriptstyle\pm}})$. It follows from these definitions that all other velocities in the problem are rescaled by $c$, $\tilde{\mathbf{v}}_{{\scriptscriptstyle\pm}}= \mathbf{v}_{{\scriptscriptstyle\pm}}/c$, stream functions go as $\tilde{\psi}_{{\scriptscriptstyle\pm}}= k\psi_{{\scriptscriptstyle\pm}}/c$, the curvature $\tilde{\kappa}= \kappa/k$, and stresses $\tilde{\boldsymbol\sigma}_{{\scriptscriptstyle\pm}}=\boldsymbol\sigma_{{\scriptscriptstyle\pm}}/\left(c\,k\,\mu_{{\scriptscriptstyle\pm}}\right)$. It is also convenient to define a coordinate variable in which the wave crests do not move, $\zeta\equiv\tilde{x}-\tilde{t}$. This last definition enables us to treat space and time variations in a unified way, such that $\partial/\partial \tilde{x}=\partial/\partial\zeta$ and $\partial/\partial\tilde{t}=-\partial/\partial\zeta$. For what follows, we shall drop the use of the symbol $\tilde{\,\,\,\,\,}$ and it is understood that all quantities are dimensionless, unless otherwise stated.

\section{$\mathcal{O}(b^1)$ results}
\label{SIB}

This section is devoted to $\mathcal{O}(b^1)$ calculations. Using the general solution (\ref{eq:SolBiharmonic}), we expand the equations (\ref{eq:BCSwimmerD}), (\ref{eq:ForceBalanceInterfaceD}), and (\ref{eq:AverageForceSheetD}), in order to find the following coefficients
\begin{subequations}
	\label{eq:CoeffOb1}
		\begin{align}
			\label{eq:CoeffOb1-A-1}
				A^{(1,1)}_{{\scriptscriptstyle-}}&=-\frac{\cosh H+\mu_r\sinh H}{\alpha_0}\left(1-H\,B^{(1,1)}_{{\scriptscriptstyle-}}\right)\\
			\label{eq:CoeffOb1-C-1}
				C^{(1,1)}_{{\scriptscriptstyle-}}&=\frac{H\,G^{(1,1)}_{{\scriptscriptstyle-}}}{1-H\,B^{(1,1)}_{{\scriptscriptstyle-}}}A^{(1,1)}_{{\scriptscriptstyle-}}\\
				D^{(1,1)}_{{\scriptscriptstyle-}}&=-G^{(1,1)}_{{\scriptscriptstyle-}},\quad
				E^{(1,1)}_{{\scriptscriptstyle-}}=-B^{(1,1)}_{{\scriptscriptstyle-}},\quad
				F^{(1,1)}_{{\scriptscriptstyle-}}=1,\quad
				H^{(1,1)}_{{\scriptscriptstyle-}}=0\\
				A^{(1,1)}_{{\scriptscriptstyle+}}&=\frac{\mbox{e}^H}{\alpha_0}\left[\sinh H\left[\left(\mu_r+B^{(1,1)}_{{\scriptscriptstyle-}}\right)\cosh H+\left(1+\mu_rB^{(1,1)}_{{\scriptscriptstyle-}}\right)\sinh H\right]\right.\nonumber\\
				\label{eq:CoeffOb1-A-2}
				&\left.+\left(1+H\left(\mu_r-1\right)\right)\left(1-H\,B^{(1,1)}_{{\scriptscriptstyle-}}\right)\right]\\
				B^{(1,1)}_{{\scriptscriptstyle+}}&=\frac{\mbox{e}^H}{1+H}\left[\left(1-HB^{(1,1)}_{{\scriptscriptstyle-}}\right)\cosh H+\left(H-B^{(1,1)}_{{\scriptscriptstyle-}}\right)\sinh H\right.\nonumber\\
				\label{eq:CoeffOb1-B-2}
				&\left.-\frac{H^3\left(\mu_r-1\right)\left(1-HB^{(1,1)}_{{\scriptscriptstyle-}}\right)}{\alpha_0}\right]\\
			\label{eq:CoeffOb1-C-2}
				C^{(1,1)}_{{\scriptscriptstyle+}}&=\mbox{e}^H\frac{\left(1+H\left(\mu_r-1\right)\right)H-\sinh H\left(\cosh H+\mu_r\sinh H\right)}{\alpha_0}G^{(1,1)}_{{\scriptscriptstyle-}}\\
			\label{eq:CoeffOb1-D-2}
				D^{(1,1)}_{{\scriptscriptstyle+}}&=C^{(1,1)}_{{\scriptscriptstyle+}}+\mbox{e}^HH\frac{\left(\mu_r-1\right)-\sinh H\left(\cosh H+\mu_r\sinh H\right)}{\alpha_0}G^{(1,1)}_{{\scriptscriptstyle-}}\\
			\label{eq:CoeffOb1-u}
				u^{(1,1)}_{{\scriptscriptstyle-}}&=0,\quad u^{(2,1)}_{{\scriptscriptstyle-}}=0,\quad u^{(3,1)}_{{\scriptscriptstyle-}}=0,\quad u^{(1,1)}_{{\scriptscriptstyle+}}=0,\quad\mbox{and}\quad U_{\mbox{{\tiny M}}}^{(1)}=0,
		\end{align}
\end{subequations}
where it is convenient to define the constant $\alpha_0\equiv (1+H\mu_r) \cosh H+(H+\mu_r) \sinh H$. The reason why we have not yet solved for $B^{(1,1)}_{{\scriptscriptstyle-}}$ and $G^{(1,1)}_{{\scriptscriptstyle-}}$ is because the shape of the membrane is still unknown. By looking at the Eq.~(\ref{eq:BCInterfaceD2}), written as $\left.\partial\psi_{{\scriptscriptstyle\pm}}/\partial\zeta\right|_{x_{\mbox{{\tiny M}}},y_{\mbox{{\tiny M}}}}=\partial y_{\mbox{{\tiny M}}}/\partial\zeta$, it is clear that the explicit form of $y_{\mbox{{\tiny M}}}$ is required information to complete the list of coefficients. Therefore, the shape of the membrane to first order, given by $X^{(1)}(\zeta)$ and $Y^{(1)}(\zeta)$, is found by solving (\ref{eq:BCInterfaceD1}) and (\ref{eq:ForceBalanceInterfaceD2}), which can be written to first order in $b$ as follows:
\begin{subequations}
	\label{eq:Shape}
		\begin{align}
			\label{eq:ShapeX}
				0&=\frac{\partial X^{(1)}}{\partial \zeta}+\left.\frac{\partial\Psi^{(1)}_{{\scriptscriptstyle\pm}}}{\partial y}\right|_{\zeta,H}\\
			\label{eq:ShapeY}
				0&=\left[P^{(1)}_{{\scriptscriptstyle-}}+\left.2\frac{\partial^2\Psi^{(1)}_{{\scriptscriptstyle-}}}{\partial \zeta\partial y}-\mu_r\left(P^{(1)}_{{\scriptscriptstyle+}}+2\frac{\partial^2\Psi^{(1)}_{{\scriptscriptstyle+}}}{\partial \zeta\partial y}\right)\right]\right|_{\zeta,H}+M\!a\frac{\partial^4 Y^{(1)}}{\partial \zeta^4}-\frac{1}{C\!a}\frac{\partial^2 Y^{(1)}}{\partial \zeta^2},
		\end{align}
\end{subequations}
where $X^{(1)}$ and $Y^{(1)}$ are periodic functions. Before solving the Eq.~(\ref{eq:ShapeY}), we also need the pressures $P^{(1)}_{{\scriptscriptstyle-}}$ and $P^{(1)}_{{\scriptscriptstyle+}}$, which are obtained by integrating the Stokes' equations just once and requiring that these are also periodic functions. We may now solve the equations (\ref{eq:Shape}), resulting in the following forms:   
\begin{subequations}
	\label{eq:ShapeSol}
		\begin{align}
			\label{eq:ShapeSolX}
				&X^{(1)}\!=\!\sin\zeta\left((H-1)A^{(1,1)}_{{\scriptscriptstyle+}}+B^{(1,1)}_{{\scriptscriptstyle+}}\right)\mbox{e}^{-H}-\cos\zeta\left((H-1)C^{(1,1)}_{{\scriptscriptstyle+}}+D^{(1,1)}_{{\scriptscriptstyle+}}\right)\mbox{e}^{-H}\\
				&Y^{(1)}\!=\!\frac{2C\!a}{1+M\!a\,C\!a}\nonumber\\
				&\!\!\left[\sin\zeta\left(\mu_r\left(HC^{(1,1)}_{{\scriptscriptstyle+}}+D^{(1,1)}_{{\scriptscriptstyle+}}\right)\mbox{e}^{-H}\!\!+\!\!\left(HC^{(1,1)}_{{\scriptscriptstyle-}}-G^{(1,1)}_{{\scriptscriptstyle-}}\right)\!\cosh H\!+\!G^{(1,1)}_{{\scriptscriptstyle-}}H\sinh H\right)\right.\nonumber\\
				\label{eq:ShapeSolY}
				&\!\!\!\!\left.-\cos\zeta\left(\mu_r\left(HA^{(1,1)}_{{\scriptscriptstyle+}}+B^{(1,1)}_{{\scriptscriptstyle+}}\right)\mbox{e}^{-H}\!\!+\!\!\left(HA^{(1,1)}_{{\scriptscriptstyle-}}-B^{(1,1)}_{{\scriptscriptstyle-}}\right)\!\cosh H\!+\!\left(1-HB^{(1,1)}_{{\scriptscriptstyle-}}\right)\sinh H\right)\right].
		\end{align}
\end{subequations}
Finally, we substitute the result (\ref{eq:ShapeSolY}) into $\left.\partial\psi_{{\scriptscriptstyle\pm}}/\partial\zeta\right|_{x_{\mbox{{\tiny M}}},y_{\mbox{{\tiny M}}}}=\partial y_{\mbox{{\tiny M}}}/\partial\zeta$ to solve for $B^{(1,1)}_{{\scriptscriptstyle-}}$ and $G^{(1,1)}_{{\scriptscriptstyle-}}$. This yields the following result
\begin{subequations}
	\label{eq:CoeffOb2}
		\begin{align}
			\label{eq:CoeffOb2-A-1}
				B^{(1,1)}_{{\scriptscriptstyle-}}&= \frac{2 H \left(\mu_r^2-1\right)+\left(\mu_r^2+1\right)\sinh (2 H)+2 \mu_r \cosh (2 H)}{\alpha_1} +\frac{4 \alpha_0^2 \alpha_2 (1+C\!a M\!a)^2}{\alpha_1 |\alpha_3|^2}\\
			\label{eq:CoeffOb2-C-1}
				G^{(1,1)}_{{\scriptscriptstyle-}}&= -\frac{8 \alpha_0^2C\!a (1+C\!a M\!a)}{|\alpha_3 |^2},
		\end{align}
\end{subequations}
where
\begin{subequations}
	\label{eq:CoeffConst}
		\begin{align}
			\alpha_1 &\equiv \left(2 H^2+1\right) \left(\mu_r^2-1\right)-\left(\mu_r^2+1\right) \cosh (2 H)-2 \mu_r \sinh (2 H)\\
			\alpha_2 &\equiv -\mu_r-2 H (1+H \mu_r)+\mu_r\cosh (2 H)+\sinh (2 H)\\
			\alpha_3 &\equiv(1+M\!a C\!a) \alpha_2-(1-C\!a)\sinh (2 H)+2 iC\!a\alpha_1.
		\end{align}
\end{subequations}

\section{$\mathcal{O}(b^2)$ results}
\label{SIC}

In order to find the $\mathcal{O}(b^2)$ relevant information, namely the speeds that arise in the problem, we impose the following conditions: (i) average of the non-slip along the $\zeta$-coordinate (\ref{eq:BCSwimmerD1}); (ii) the tangent force balance on the membrane (\ref{eq:ForceBalanceInterfaceD3}); (iii) force free condition on the swimmer (\ref{eq:AverageForceSheetD}); and the definition of the interface's average speed in the swimmer's frame, which can be split into two conditions, (iv) below ($-$) and (v) above ($+$) the membrane. Therefore, the conditions from (i) to (v) are enough to determine the five unknown coefficients $\{u^{(1,2)}_{{\scriptscriptstyle-}},u^{(2,2)}_{{\scriptscriptstyle-}},u^{(3,2)}_{{\scriptscriptstyle-}},u^{(1,2)}_{{\scriptscriptstyle+}},U_{\mbox{\tiny I}}^{(2)}\}$:
\begin{subequations}
	\label{eq:Speeds}
		\begin{align}
			\label{eq:Speed1}
				u^{(1,2)}_{{\scriptscriptstyle-}}& = -\frac{1}{2} +\frac{(1-H B_{{\scriptscriptstyle-}}^{(1,1)}) (\cosh  H +\mu_r \sinh  H )}{\alpha_0 }\\
			\label{eq:Speed2}
				u^{(2,2)}_{{\scriptscriptstyle-}}& = 0\\
			\label{eq:Speed3}
				u^{(3,2)}_{{\scriptscriptstyle-}}& = 0\\
			\label{eq:Speed4}
				u^{(1,2)}_{{\scriptscriptstyle+}}& = -\frac{1}{2}+\frac{(\cosh  H +\mu_r \sinh  H ) \left[(1+M\!a  C\!a) (1-H B_{{\scriptscriptstyle-}}^{(1,1)})+2C\!a  H (1-\mu_r) G_{{\scriptscriptstyle-}}^{(1,1)}\right]}{(1+M\!a  C\!a ) \alpha_0}\\
				U_{\mbox{\tiny I}}^{(2)}& = -\frac{1}{2}+\frac{(\cosh  H +\mu_r \sinh  H )\left[(1+M\!a  C\!a) (1-HB_{{\scriptscriptstyle-}}^{(1,1)})-2C\!a H\mu_rG_{{\scriptscriptstyle-}}^{(1,1)}\right]}{(1+M\!a  C\!a ) \alpha_0}\nonumber\\
				    			\label{eq:Speed5}
				&+\frac{C\!a G_{{\scriptscriptstyle-}}^{(1,1)}}{1+M\!a  C\!a}.
		\end{align}
\end{subequations}

\section{Validation of results}
\label{SID}

From our approach we are able to verify results that have been previously treated in the literature. Here, we consider two limiting cases, the results for swimming in an unbounded fluid \cite{Taylor1951}, hence no-membrane limit, and swimming near a rigid-wall \cite{Reynolds1965,Katz1974}. The former is derived by taking the limit where ${M\!a}\rightarrow0$, ${C\!a}\rightarrow\infty$, and $\mu_r\rightarrow1$, while the latter is recovered in the limit where ${M\!a}\rightarrow\infty$, ${C\!a}\rightarrow0$, and $\mu_r\rightarrow\infty$. The first order stream functions, for the respective limits, are given by
\begin{widetext}
\begin{subequations}
	\label{eq:LimitStreamFunc}
		\begin{align}
			\label{eq:LimitStreamFunc1}
\lim_{M\!a\rightarrow 0,\,Ca\rightarrow \infty,\,\mu_r\rightarrow1}\Psi^{(1)}_{{\scriptscriptstyle-}}(\zeta,y)&=\lim_{M\!a\rightarrow 0,\,Ca\rightarrow \infty,\,\mu_r\rightarrow1}\Psi^{(1)}_{{\scriptscriptstyle+}}(\zeta,y)= e^{-y} (1+y) \sin\zeta\\
				\lim_{M\!a\rightarrow\infty,\,Ca\rightarrow 0,\,\mu_r\rightarrow\infty}\Psi^{(1)}_{{\scriptscriptstyle-}}(\zeta,y)&=\sin\zeta\left[\cosh  y \left( 1- y\frac{H+\cosh  H\sinh  H}{ H^2-\sinh ^2 H}\right)\right.\nonumber\\
			\label{eq:LimitStreamFunc2}
				&\quad\quad\quad\left.+\sinh  y \left(\frac{H+\cosh  H \sinh H}{ H^2-\sinh ^2 H}+ y \frac{\sinh ^2 H}{ H^2-\sinh ^2 H}\right)\right]\\
			\label{eq:LimitStreamFunc3}
				\lim_{M\!a\rightarrow\infty,\,Ca\rightarrow 0,\,\mu_r\rightarrow\infty}\Psi^{(1)}_{{\scriptscriptstyle+}}(\zeta,y)&=0.
\end{align}
\end{subequations}
\end{widetext}
In the no-membrane limit, the average height and the displacement of fluid material points take the form $Y^{(1)}=(1+H)\,\mbox{e}^{-H}\sin\zeta$ and $X^{(1)}=-H\,\mbox{e}^{-H}\cos\zeta$, respectively. As expected, in the rigid-wall limit we have $Y^{(1)}=X^{(1)}=0$. In contrast with the results found in the lubricating limit for a free-interface ($\mu_r=0$ and $M\!a\rightarrow0$) \cite{Lee2008}, we generalize this result by taking the limit ${M\!a}\rightarrow0$, which yields expressions for a free-interface at finite height and viscosity mismatch.

\end{document}